\documentclass[lettersize,journal]{IEEEtran}
\usepackage{graphicx}
\usepackage{amssymb}
\usepackage{amsmath}
\usepackage{mathtools}
\usepackage{cite}
\usepackage{stfloats}
\usepackage{epstopdf}
\usepackage{psfrag}
\usepackage[mathscr]{euscript}
\usepackage{acronym}  % make an acronym
\usepackage{booktabs}
\usepackage[table]{xcolor}
\usepackage{tikz}
\usepackage{url}
\usepackage[many]{tcolorbox}
\usepackage{algorithm}
\usepackage[noend]{algpseudocode}
\usepackage{kotex}
\usepackage{float}
\def\BibTeX{{\rm B\kern-.05em{\sc i\kern-.025em b}\kern-.08em
		T\kern-.1667em\lower.7ex\hbox{E}\kern-.125emX}}
\usepackage[breaklinks, plainpages=false, pdfpagelabels=false, bookmarksnumbered=false]{hyperref} % create hyperlinks
\usepackage{breakurl}

\usepackage{color}
\usepackage{dsfont}
\usepackage{bbm}

\newcommand{\Ri}[1]{r_{#1}}

\newcommand{\AngToDist}[1]{\frac{180}{\pi}\text{arctan}\left(\frac{\ell}{\Ri{i}}\right)}

\renewcommand\IEEEkeywordsname{Keywords}

\newcommand{\LLMhRIC}{LLM-\emph{h}RIC}

\acrodef{MBS}{macro base station}
\acrodef{SBS}{small base station}
\acrodef{OFDMA}{orthogonal frequency-division multiple access}
\acrodef{LoS}{line-of-sight}
\acrodef{NLoS}{non-line-of-sight}
\acrodef{AWGN}{additive white Gaussian noise}
\acrodef{SINR}{signal-to-interference-plus-noise ratio }
\acrodef{PPO}{proximal policy optimization}
\acrodef{MARL}{multi-agent reinforcement learning}
\acrodef{DDPG}{deep deterministic policy gradient}
\acrodef{MDP}{markov decision process}
\acrodef{DRL}{deep reinforcement learning}
\acrodef{LLM}{large language model}
\acrodef{SLM}{small language model}
\acrodef{O-RAN}{open radio access network}
\acrodef{RAN}{radio access network}
\acrodef{RIC}{\ac{RAN} intelligent controller}
\acrodef{LLM-hRIC}{LLM-empowered hierarchical RIC}
\acrodef{RU}{radio unit}
\acrodef{CU}{central unit}
\acrodef{DU}{distributed unit}
\acrodef{near-RT}{near-real-time}
\acrodef{non-RT}{non-real-time}
\acrodef{near-RT RIC}{near-real-time \ac{RIC}}
\acrodef{non-RT RIC}{non-real-time \ac{RIC}}
\acrodef{CNN}{conventional neural network}
\acrodef{IAB}{integrated access and backhaul}
\acrodef{gNB}{next generation node B}
\acrodef{RL}{reinforcement learning}
\acrodef{ML}{machine learning}
\acrodef{RAG}{retrieval augmentd generation}
\acrodef{PRB}{physical resource block}
\acrodef{SMO}{service management and orchestration}
\begin{document}

\newcommand{\paperTitle}{
{\LLMhRIC}: LLM-empowered Hierarchical RAN Intelligent Control for O-RAN \\
}

%---------------------------------------------------------------------------%
%                     title, title footnote, header                         %
%---------------------------------------------------------------------------%

\title{\paperTitle}

\author{Lingyan Bao, Sinwoong Yun, Jemin Lee, and Tony Q.S. Quek
\thanks{
	Jemin Lee (corresponding author) and Lingyan Bao are with the School of Electrical and Electronic Engineering, Yonsei University, South Korea; 
	
	Sinwoong Yun is with the ICT Strategy Research Labortory, Electronics and Telecommunications Research Institute (ETRI), South Korea; 
	
	Tony Q. S. Quek is with the Information System Technology and Design, Singapore University of Technology and Design, Singapore. 

    This work was supported in part by the Institute of Information \& communications Technology Planning \& Evaluation (IITP) grants funded by the Korea government (MSIT) (No. 2024-00404972, Development of 5G-A vRAN Research Platform; in part by the IITP under 6G·Cloud Research and Education Open Hub (IITP-2025-RS-2024-00428780) grant funded by the Korea government (MSIT); and in part by the 6GARROW project, which has received funding from the Smart Networks and Services Joint Undertaking (SNS JU) under the European Union’s Horizon Europe research and innovation programme under Grant Agreement No 101192194 and from the Institute for IITP grant funded by the Korean government (MSIT) (No. RS-2024-00435652).
}
}

\maketitle %% make the title area
\setcounter{page}{1}
\renewcommand\IEEEkeywordsname{Index Terms}

%%---------------------------------------------------------------------------%
%%                           abstract and key words                          %
%%---------------------------------------------------------------------------%
\begin{abstract}
Despite recent advances in applying \acp{LLM} and \ac{ML} techniques to \ac{O-RAN}, critical challenges remain, such as insufficient cooperation between \acp{RIC}, high computational demands hindering real-time decisions, and the lack of domain-specific fine-tuning.
Therefore, this article introduces the LLM-empowered hierarchical RIC ({\LLMhRIC}) framework to improve the collaboration between \acp{RIC} in \ac{O-RAN}. 
The LLM-empowered \ac{non-RT RIC} acts as a \emph{guider}, offering a strategic guidance to the \ac{near-RT RIC} using global network information. The RL-empowered \ac{near-RT RIC} acts as an \emph{implementer}, combining this guidance with local real-time data to make near-RT decisions.
We evaluate the feasibility and performance of the {\LLMhRIC} framework in an \ac{IAB} network setting,
and finally, discuss the open challenges of the {\LLMhRIC} framework for \ac{O-RAN}. 
\end{abstract}


\acresetall
\section{Introduction}
The \ac{O-RAN} has recently gained significant attention for its ability to prompt interoperability and flexibility. 
In \ac{O-RAN}, the \ac{gNB} is functionally split into three components: the \ac{RU}, the \ac{DU}, and the \ac{CU}. The different components of the networks can be developed and supplied by various vendors through open standards and interfaces. This openness enables a competitive ecosystem, driving innovation and reducing costs. 
Moreover, by introducing a modular, software-driven architecture, \ac{O-RAN} transforms the traditional hardware-centric approach to virtual \ac{RAN} \cite{polese2023understanding}. 

A core component of \ac{O-RAN} is the \acp{RIC}, which provides advanced control and optimization capabilities. The \acp{RIC} include: 1) \emph{the \ac{non-RT RIC}}, which integrates with the network orchestrator and operates on a time scale larger than 1 s, and 2) \emph{the \ac{near-RT RIC}}, which manages control loops with \ac{RAN} nodes on a time scale between $10$ ms and $1$ s. 
The \ac{non-RT RIC} and the \ac{near-RT RIC} are capable of obtaining key performance measurements (KPMs) via O1 and E2 interfaces, respectively, to monitor the status of network infrastructure (e.g., number of users, load, throughput, and resource utilization).  
The \ac{non-RT RIC} also supports the deployment of policies, guidance, and intelligent models within the \ac{near-RT RIC} via the A1 interface.
Specific RIC functionalities, such as mobility management and network slicing, are deployed within \ac{near-RT RIC} and \ac{non-RT RIC} in the form of the third-party applications, known as xApp and rApp, respectively \cite{abdalla2022toward}.

Recently, intensive researches on \acp{RIC} have been appeared\cite{liang2024enhancing,nguyen2023network,d2023orchestran,amiri2023deep,zhang2023learning,filali2023communication,sohaib2025optimizing}. 
Specifically, the \acp{near-RT RIC} has been developed to manage a smaller number of devices and optimize on a small-time scale, such as the radio card switching for energy efficiency \cite{liang2024enhancing} and the flow-split distribution, congestion control, and scheduling for higher network utility \cite{nguyen2023network}. In particular, \ac{DRL} has emerged as a key technology for intelligent xApps, enabling dynamic control of radio resource management in \ac{O-RAN} \cite{sohaib2025optimizing}.
The \ac{non-RT RIC} has been developed to manage a larger number of devices and perform high-level and long-term network management, such as the orchestRAN\cite{d2023orchestran}, which automatically computes the optimal set of algorithms and their execution location.
Furthermore, the cooperation between \ac{non-RT RIC} and \ac{near-RT RIC} is investigated in  \cite{amiri2023deep,zhang2023learning,filali2023communication}. Specifically, the \ac{non-RT RIC} provides the high-level guidance to \ac{near-RT RIC}, including the network configuration and reconfiguration information \cite{amiri2023deep}, control signals \cite{zhang2023learning}, and optimized control functions \cite{filali2023communication}. The \ac{near-RT RIC} then refines and implements these policies based on real-time information to enhance the network performance. 

Meanwhile, the \acp{LLM} have opened new routes for wireless network design and operation including O-RAN.
The \acp{LLM} are advanced \ac{ML} models, designed to understand and generate human language. Their ability to understand context and generate contextually relevant outputs makes them highly adaptable across various domains.
In the context of telecommunications, \acp{LLM} offer unique advantages by processing and analyzing extensive network data to improve decision-making. 
Recent studies have applied \ac{LLM}-based technologies to wireless network 
for the resource allocation\cite{lee2024llm,zhou2024large}, 
base station placement\cite{wang2024large}, 
and packet analysis \cite{kan2024mobile}.
Recently, the \ac{LLM}-based xApp to allocate \ac{PRB} for \ac{O-RAN} has also been introduced\cite{wu2025llm}.

Despite advancements in \acp{RIC} and \ac{LLM}-based techniques for wireless networks, a critical real-world problem remains: bridging the `coordination gap' between the strategic, long-term control of the \ac{non-RT RIC} and the tactical, real-time execution of the \ac{near-RT RIC}. This gap is exacerbated by several underlying challenges:
1) traditional \ac{ML} techniques struggle to process the diverse and complex information in \ac{O-RAN}; 2) the efficient cooperation between \ac{non-RT RIC} and \ac{near-RT RIC} still remains largely unexplored; 3) although the potential on the role of \acp{LLM} in wireless networks has been shown, the high computational complexity hinders near-RT decision-making; and 4) without fine-tuning for domain-specific tasks, the performance of \ac{LLM}-based techniques often remains comparable to or worse than traditional algorithms.

Motivated by these opportunities and challenges, we propose the novel framework for \ac{O-RAN}, which is the  {\LLMhRIC}. In this framework, there are two \acp{RIC}: 1) the \acp{LLM}-empowered \ac{non-RT RIC} and 2) the \ac{RL}-empowered \ac{near-RT RIC}. The \acp{LLM}-empowered \ac{non-RT RIC} works as a \emph{guider} to the {near-RT RIC}. Based on the objective of the \ac{near-RT RIC}, it provides strategic guidance for the \ac{RL}-empowered \ac{near-RT RIC} based on \emph{the global view of the networks} (e.g., long-term channel information, network nodes state, users mobility model, and users distribution).
The \ac{RL}-empowered \ac{near-RT RIC} works as an \emph{implementer}, which utilizes this guidance along with \emph{the local view of the networks} (e.g., real-time channel quality and local network node state) to perform near-RT decision-making. 
The key contribution of the paper can be summarized as follows. 

\begin{itemize}
	\item We propose the novel {\LLMhRIC} framework for O-RAN, which can show how the LLM can be utilized to enhance the performance of RICs in O-RAN. 
	This article focuses on addressing the limitation of LLM in making near-RT decisions and paves a way to utilize it in a reliable way in O-RAN environments.  	
	\item We introduce the three sequential phases for training of the \ac{RL}-empowered \ac{near-RT RIC}, of each phase utilizes the guidance of the \ac{non-RT RIC} in a different way to accelerate convergence as well as improve the system performance. 
	\item We quantify the performance of the {\LLMhRIC} for the power allocation in the \ac{IAB} networks as a use case. This can demonstrate the ability of {\LLMhRIC} to manage the complex information in \ac{O-RAN}. 
\end{itemize}

\section{LLM-empowerd hierarchical RIC Framework}\label{sec:3}
\begin{figure*}[t]
	\centering\includegraphics[width=0.83\linewidth]{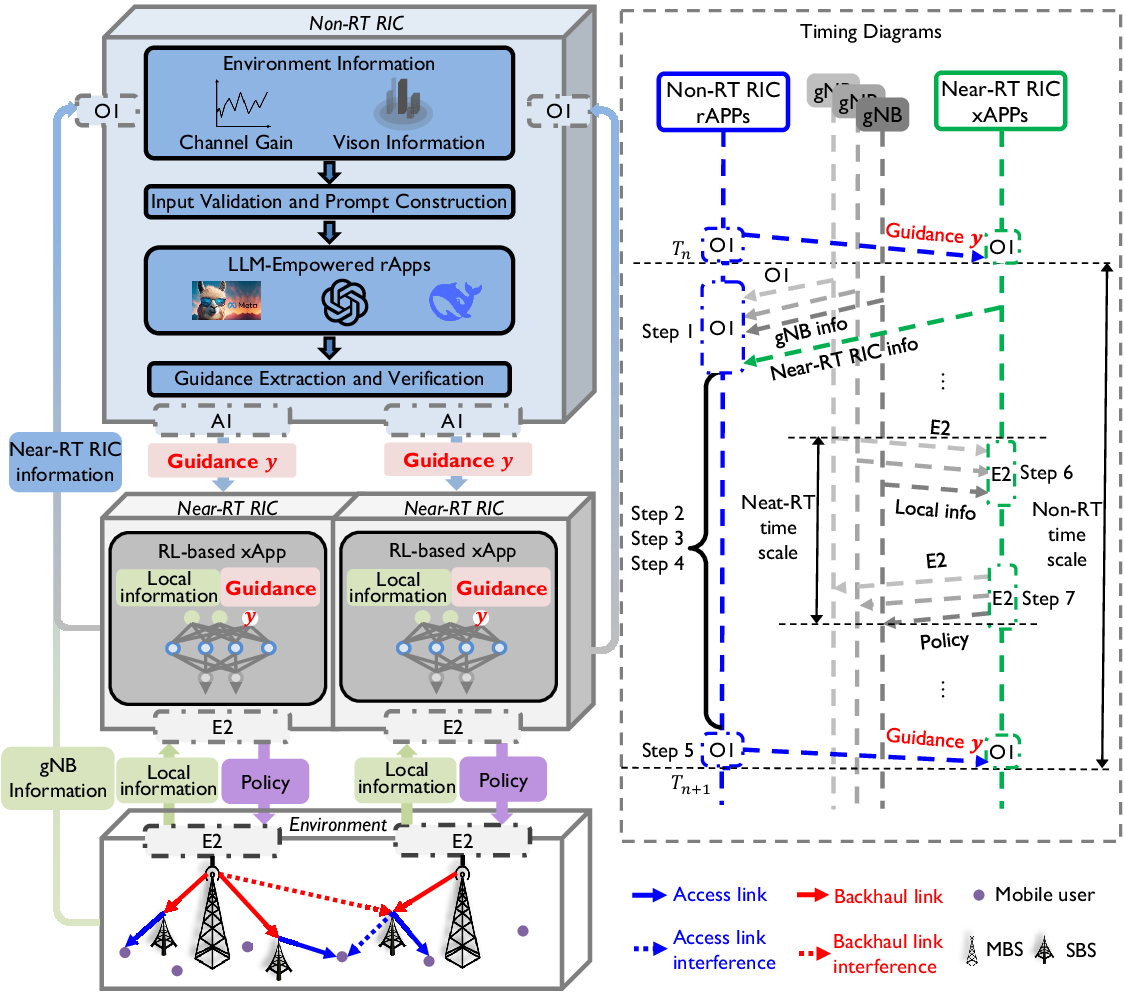}
	\caption{Structure of the proposed {\LLMhRIC} framework and time diagrame.}
	\label{fig:HLR-RIC}
\end{figure*}

In this section, we introduce the LLM-empowered hierarchical RIC ({\LLMhRIC}) framework for resource management in O-RAN as shown in Fig. \ref{fig:HLR-RIC}. The {\LLMhRIC} framework consists of 1) the LLM-empowered non-RT RIC (i.e., upper layer control) and 2) the RL-empowered near-RT RIC (i.e., lower layer control), and this shows how the LLM can be used as a way to facilitate the resource management in O-RAN. 
As the LLM has the advantages of generalization across various environments with minimal language data, it is used for the non-RT RIC to provide guidance on the strategies to the \ac{near-RT RIC} from the large-scale view of the networks. The \ac{near-RT RIC} can be designed to utilize this guidance in conjunction with the small-scale view to make decisions in near-RT. 
More details on the {\LLMhRIC}  framework are provided in the following subsections. 

\vspace{-2mm}
\subsection{LLM-Empowered \ac{non-RT RIC}}
The \ac{non-RT RIC} serves as a strategic layer, which works based on the collected KPMs from various network units including \ac{gNB} and \ac{near-RT RIC} via O1 interface. 
The rApps, deployed for the \ac{non-RT RIC}, are responsible for utilizing that information to generate strategic guidance for the \ac{near-RT RIC}.
By integrating \ac{LLM} technology, the \ac{non-RT RIC} gains the capability to process and analyze multi-modal data to form a comprehensive understanding of the current network environments.
Additionally, \acp{LLM} excels in synthesizing diverse data sources and leveraging their extensive parameters and deep architecture to perform complex reasoning and inference. These strengths allow the \ac{LLM}-empowered \ac{non-RT RIC} to tackle challenges in complex network management and provide high-quality strategic guidance to the \ac{near-RT RIC}. 

The process of the LLM-empowered \ac{non-RT RIC} can be summarized as following steps. 

\begin{itemize}[]
	\item[1)] \emph{Data Integration}: 
        The non-RT RIC ingests multi-modal data from \ac{gNB} and \ac{near-RT RIC}, primarily via the O1 interface. These data include long-term time-series metrics (e.g., total throughput trends), unstructured textual data (e.g., fault alarm logs), topology data representing network node connectivity, and even advanced visual data such as radio maps generated from digital twins.
    \item[2)] \emph{Input Validation and Prompt Construction}:
        Before being formatted into a prompt, this data undergoes a validation stage to handle potentially noisy or corrupted measurements. 
        The data integration process involves structuring these varied inputs into a unified, context-rich format that can be effectively processed by the \ac{LLM}. 
	\item[3)] \emph{Environment Understanding and Actionable Insights Generation}: 
	The \ac{LLM} analyzes and understands the context based on the integrated multimodal data, and 
	generate the text response, that includes the strategic guidance $\textbf{y}$ (e.g., transmit power allocation strategies).
    \item[4)] \emph{Guidance Extraction and Verification}: The LLM's raw text output is validated for its conformance to predefined formats and constraints, and extracts the strategic guidance $\textbf{y}$. To ensure robustness, if this process fails, a fallback mechanism provides a default policy as guidance.
	\item[5)] \emph{Guidance Transfer to the \ac{near-RT RIC}}: 
	The guidance vector $\textbf{y}$ is then sent to the \ac{near-RT RIC} via the A1 interface, used for the near-RT decision-making process.
\end{itemize}

\subsection{RL-Empowered \ac{near-RT RIC}}
Generally, given the strict latency constraints of the \ac{near-RT RIC} (e.g., 10 ms to 1s), the xApps need to be designed to process information efficiently and make fast decisions in a highly dynamic network environment. Hence, we employ \ac{RL} because its ability to learn an adaptive policy is well-suited for reacting to the dynamic environments with real-time changes. Furthermore, RL is naturally designed to incorporate a strategic guidance vector as a part of its state input, fusing it with the observed local information to make refined and context-aware decisions.

The process of the RL-empowered \ac{near-RT RIC} can be summarized in the following steps: 
\begin{itemize}[]
	\item[6)] \emph{Data Integration}: 
    The \ac{near-RT RIC} ingests the strategic guidance from the non-RT RIC (via the A1 interface) and the observed local information from the \ac{gNB} (via the E2 interface).
	\item[7)] \emph{Policy generation}: 
	The \ac{RL}-based xApp then analyzes these inputs to generate an optimal, fine-grained policy that is sent to the \ac{gNB} for execution.
\end{itemize}

This integration of global insights with real-time local information enhances the RL agent's ability to make refined, fine-grained decisions that are both strategically sound and tactically optimal for the immediate radio conditions.

\section{Use Case: Power Allocation in \ac{IAB} Network} \label{sec:use case}
In this section, we present a use case based on \ac{IAB} networks to verify the effectiveness of the proposed {\LLMhRIC} framework. This scenario was chosen as a foundational proof-of-concept since the resource allocation in multi-cell IAB networks  
    is a well-known difficult problem to handle as it requires the network-wise global information and control.  
    This problem can demonstrate our framework's strength: using the \ac{LLM} in the non-RT RIC to provide a globally-aware and coordinated guidance to local RL agents.

\subsection{Power Allocation in \ac{IAB} Networks}
The \ac{IAB} networks as illustrated in Fig. \ref{fig:HLR-RIC}, consists of a set of \acp{MBS}, denoted as $\mathcal{M} = \{1,\cdots, M\}$, distributed in a network. Each \ac{MBS} is connected with $N$ \acp{SBS}, and $\mathcal{N}_m = \{1,\cdots,N\}$ denotes the set of \acp{SBS}, connected to the $m$-th \ac{MBS}. A set of mobile users $\mathcal{K}_{m,n}$, connected with the $n$-th \ac{SBS} of the $m$-th \ac{MBS}, follow a Gaussian-Markov mobility model. The \ac{SBS} allocates fixed power to their connected user.
The interference exists among \acp{SBS} connected to different \ac{MBS} that use the same sub-carrier. Both \ac{LoS} and \ac{NLoS} path loss characteristics are considered, where the probability of a \ac{LoS} connection is modeled as $P^{\text{Los}} = \text{exp}(-d/\rho)$, where $d$ represents the distance between \ac{MBS} and \ac{SBS}, and $\rho$ is a \ac{LoS} range constant. 
Furthermore, the small-scale fading is modeled as independent Nakagami-$m$ fading. 

In this use case, the total channel bandwidth $W$ for each \ac{MBS} is allocated between the backhaul and access links in proportion $\alpha$, where $\alpha \in [0,1]$, i.e., the fraction of bandwidth allocated to the backhaul link is $W\alpha$ and the remaining bandwidth is assigned to the access links.
At the $t$-th time slot, the achievable rate of the access link between the $k$-th user and the $n$-th SBS
and that of the backhaul link between the $m$-th MBS and the $n$-th SBS are denoted by $R_{m,n,k}^{\text{a}}[t]$ and $R_{m,n}^{\text{b}}[t]$, respectively, and then the total throughput becomes $\mathcal{T} = \sum_{t=1}^T \sum_{m=1}^M \sum_{n \in \mathcal{N}_{m}} \min\left(R_{m,n}^{\text{b}}[t],\sum_{k \in \mathcal{K}_{m,n}} R_{m,n,k}^{\text{a}}[t] \right)$. The objective is to maximize the total throughput $\mathcal{T}$ by optimizing the power allocation of \acp{MBS} to each \acp{SBS} under the total power constraint on the \ac{MBS}, given by $\sum_{n \in \mathcal{N}_{m}} P_{m,n}[t] \leq P_{m}^{\text{max}}, \forall m,t$, where $P_{m}^{\text{max}}$ is the maximum available power of the $m$-th \ac{MBS}.

\subsection{{\LLMhRIC} Framework for Power Allocation}
In this section, the proposed {\LLMhRIC} framework is deployed for power allocation in the \ac{IAB} networks. Each \ac{MBS} connects to a \ac{near-RT RIC}, while a centralized \ac{non-RT RIC} collects information from all \acp{MBS} and provides strategic power allocation guidance to the \acp{near-RT RIC}.
\subsubsection{\ac{LLM}-Empowered \ac{non-RT RIC}} 
In this framework, the \ac{LLM} is deployed in the \ac{non-RT RIC} as a rApp.  A structured prompt guides the \ac{LLM} in generating the power allocation guidance for \ac{near-RT RIC} (e.g., initial power allocation policy). This prompt includes the following components: 
\begin{itemize}[]
	\item Objective Function: Defines the goal of power allocation optimization, i.e., to maximize the total throughput.
	\item Environment Description: Details the number of \acp{MBS}, \acp{SBS}, the maximum power of each \ac{MBS}, etc,.
	\item Channel Information: Provides the average channel gain between \acp{MBS} and \acp{SBS}.
	\item Interference Information: Specifies interference sources and their average interference channel gains.
	\item Constraints: Details the constraints on power allocation policy and the guidance format.
\end{itemize}

The structure of the prompt is as Fig.~\ref{fig:prompt-structure}.
%	\centering
\begin{figure}
	\centering
	\begin{tcolorbox}[
		colback=yellow!10!white, 
		colframe=black, 
		title=Prompt Structure
		]
		You are an expert in wireless communications for resource allocation. \\
		Your objective is \textless Objective function \textgreater. \\
		Here is the system description: \textless Environment information \textgreater. \\
		Input Format: For each MBS, you are provided with the following input data for its connected SBSs. Each SBS is represented as a list: [Average channel gain, number of connected users, average expected data rate of connected users (Mb/s), and interference: [interference source (MBS, SBS), average interference channel gain] ]. \\
		Constraints: Ensure the total power allocation across SBSs for each MBS sums to 1. For each MBS, output the normalized power allocation ratios as a list of six values corresponding to its six SBSs. MBSX: [value1, value2, value3, value4, value5, value6].
	\end{tcolorbox}
	\caption{The Prompt Structure.}
	\label{fig:prompt-structure}
\end{figure}
Due to the time-varying nature of small-scale fading, the \ac{non-RT RIC} uses average values of small-scale fading parameters to compute metrics like channel gain, expected data rate, and interference. These average values effectively represent the channel's state in future time slots.

To ensure the reliability of these inputs, a data checking step performs range validation. This process verifies that all metrics are physically plausible, such as ensuring that the `number of connected users' is non-negative and that the `average channel gain' is within a realistic range. 

\subsubsection{RL-Empowered \ac{near-RT RIC}} In the \ac{IAB} networks, each \ac{MBS} is associated \ac{RL}-based xApps for near-RT resource management.
The \ac{MDP} modeled as following components:
\begin{itemize}[]
	\item State Space: We define the system state space of $m$-th \ac{MBS} as $\mathcal{S}_m=\{\boldsymbol{h}_{m},\boldsymbol{n}_m,\boldsymbol{R}_m,\boldsymbol{p}^o_m\}$, where $\boldsymbol{h}_{m}$ is the channel gain between \ac{MBS} and its connected \acp{SBS}, $\boldsymbol{n}_m$ represents the number of associated users per connected \ac{SBS}, $\boldsymbol{R}_m$ is the average expected data rate of connected users per \ac{SBS}, and $\boldsymbol{p}^o_m$ is the initial power allocation policy provided by the \ac{non-RT RIC}.
	\item Action Space: Each \ac{MBS} allocates power to its connected \acp{SBS}, denoted as $\boldsymbol{p}_m =(p_{m,1},\cdots,p_{m,N})$, where $p_{m,n} \in [0,1]$ is the power allocation ratio for each connected \ac{SBS}.
	\item Reward Function: The reward is designed to encourage individual performance and global cooperation $r_m = r^\text{l}_m + r^{\text{g}}$, where $r^\text{l}_m$ is the throughput of $m$-th \ac{MBS} and $r^{\text{g}}$ is the total throughput of all \acp{MBS}.
\end{itemize}

\vspace{-3mm}
\subsection{\ac{LLM}-assisted Three Phases Training for \ac{RL}}
The \ac{RL} agent in the \ac{near-RT RIC} must be trained to effectively utilize the guidance provided by \ac{LLM}. Traditional \ac{RL} typically relies on a random policy for action exploration during the initial stage of training. To improve the exploration efficiency, this random policy is replaced with the \ac{LLM} guidance, such as an initial power allocation policy and injected action noise. The policy is gradually formed by blending the actions, generated by the \ac{RL} agent with the \ac{LLM} guidance, until the agent is capable of making decisions independently. Specifically, the training can be processed by the following three sequential phases:
\begin{itemize}[]
	\item Phase 1 (\ac{LLM}-guided Exploration): The power allocation policy $\boldsymbol{p}_m$ is determined by adding noise to the \ac{LLM}'s initial guidance policy $\boldsymbol{p}_m^o$, i.e., $\boldsymbol{p}_m = \boldsymbol{p}_m^o + \text{noise}$. This encourages exploration in the vicinity of the LLM’s guidance, allowing the \ac{RL} agent to collect high-quality experiences and learn a near-optimal initial policy.
	\item Phase 2 (Policy Blending): The power allocation policy $\boldsymbol{p}_m$ is determined by a weighted combination of the \ac{LLM}-provided policy $\boldsymbol{p}_m^o$ and the \ac{RL}-generated policy $\boldsymbol{p}_m^d$, i.e., $\boldsymbol{p}_m = w \boldsymbol{p}_m^o + (1-w) \boldsymbol{p}_m^d$, where the blending coefficient $w$ decays gradually from $1$ to $0$ over time. This creates a smooth transition from guided exploration to self-directed optimization.
	\item Phase 3 (Self-directed Decision): The power allocation policy $\boldsymbol{p}_m$ is determined solely by the \ac{RL} agent, i.e., $\boldsymbol{p}_m = \boldsymbol{p}_m^d$.
\end{itemize}
In each training stage, the power allocation policy interacts with the environment, and the resulting \ac{MDP} experiences are collected to train the \ac{RL} agent. 
This phased approach prevents the instability seen with abrupt policy changes, allowing the agent to efficiently fine-tune its near-optimal policy, which leads to stable convergence and the highest overall throughput.

\subsection{Experimental Results}
Our simulation deploys the Llama-3.1-8B-instruct model \cite{dubey2024llama} in the rApp to generate a power allocation guidance policy on a 2s timescale. For efficient deployment, the model is loaded with bfloat16 precision, with inference parameters of temperature $=0.6$ and top\_p $=0.9$. 
Meanwhile, a \ac{DDPG} agent in the xApp provides a fine-grained power allocation policy for a \ac{MBS} on a 200ms timescale. The DDPG agent was trained using the Adam optimizer with a learning rate of $10^{-4}$ and a batch size of 256. \footnote{Note that the demonstration on a prototype testbed can be done once O1 and A1 interfaces become sufficiently mature in open-source O-RAN platforms.}
To assess the effectiveness of the proposed framework, we compare it against the following baseline methods: 1) \ac{DDPG} with linear decay noise for action exploration (DLN), 2) \ac{DDPG} with cosine decay noise for action exploration (DCN), and 3) equal power allocation (EPA).

\begin{figure}[t]
	\centering
	\includegraphics[width=0.9\linewidth]{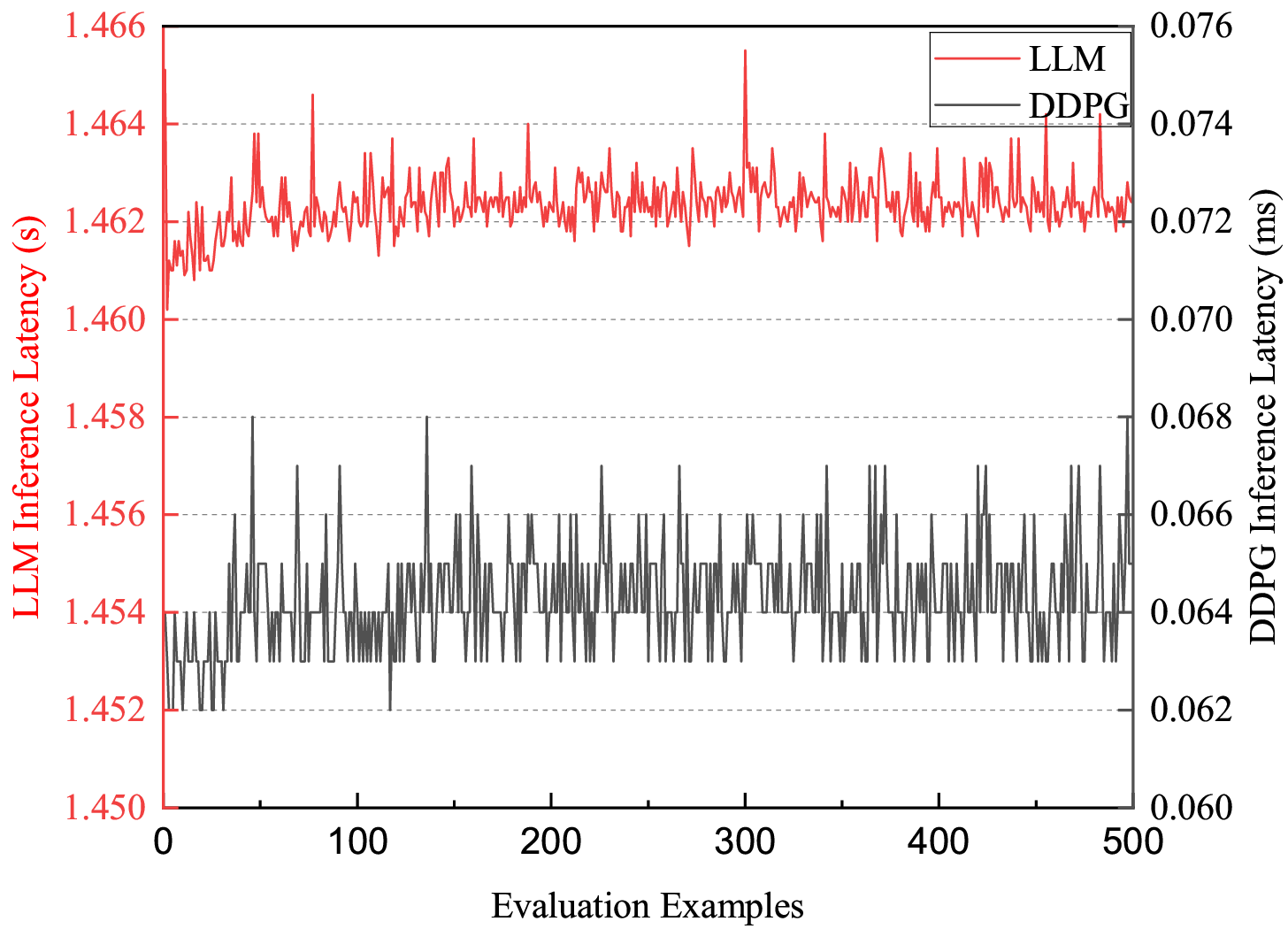}
	\caption{LLM and DDPG inference latency with 500 examples.}
	\label{fig:inference latency}
\end{figure}
Fig. \ref{fig:inference latency} shows the \ac{LLM} and \ac{DDPG} inference latency with $500$ examples. The \ac{LLM} inference latency is less than 1.5 seconds, confirming its decision time is well within the operational budget required for the {\LLMhRIC} ($\leq$ 2s). Moreover, the \ac{DDPG} inference time is less than $0.07$ ms. This extremely low latency is significantly faster than the 10 ms requirement for the most demanding near-real-time control.

\begin{figure}[t]
	\centering
    \includegraphics[width=0.8\linewidth]{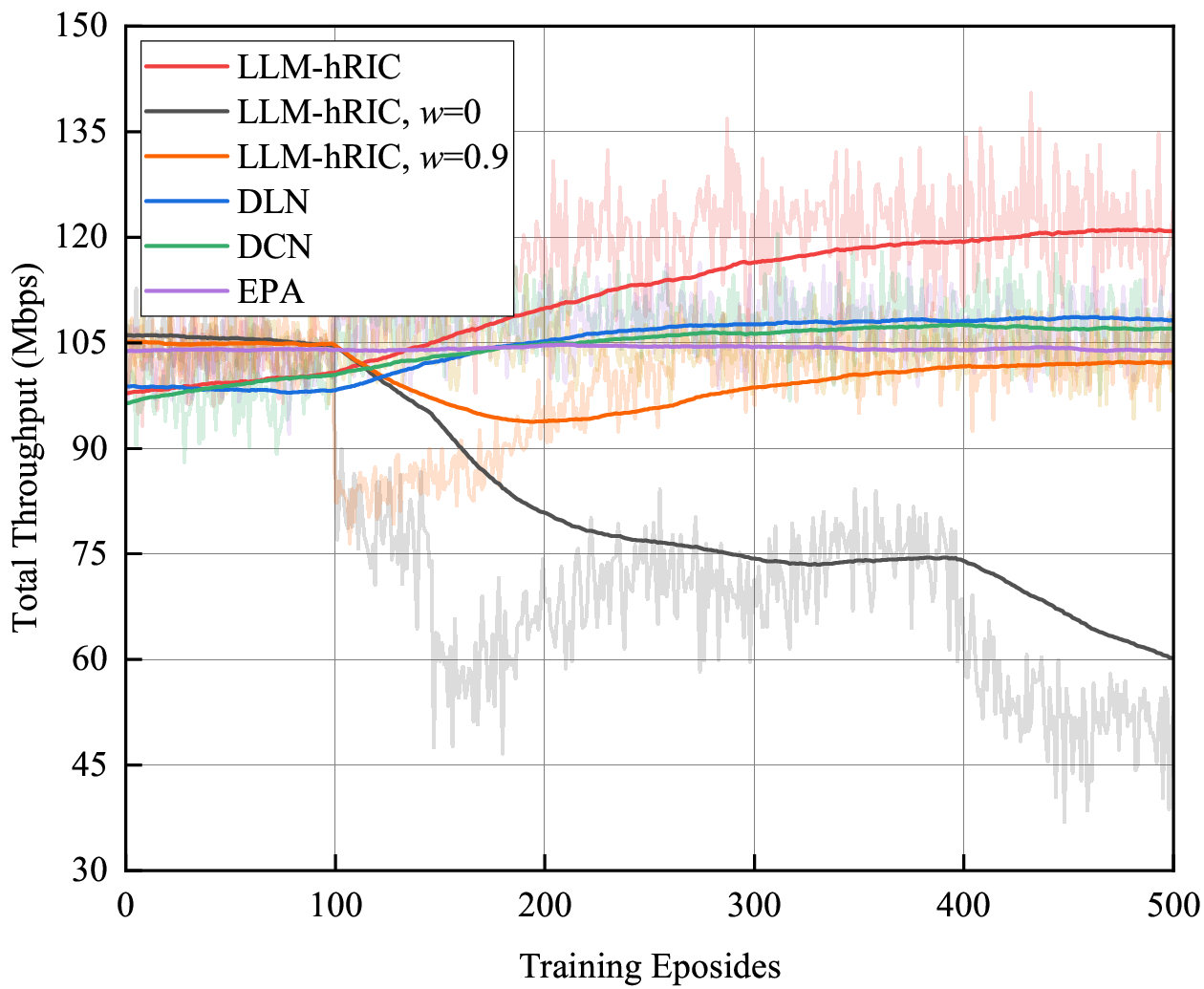}
	\caption{Training curve for proposed with different blending coefficient $w$ and baseline methods over 500 training epochs when $M=3, N=6, W=100\text{Mb}, P_m^{\text{max}}=44\text{dBm}$, and $\alpha=0.5$.}
	\label{fig:Training_epoches}
\end{figure}
Fig. \ref{fig:Training_epoches} shows the training curves of proposed {\LLMhRIC} framework and baseline methods over $500$ training epochs when the backhaul-access bandwidth partitioning ratio $\alpha=0.5$. The results show that the total throughput increases steadily as training progresses and gradually converges as the models learn effective policies. Notably, the proposed {\LLMhRIC} framework achieves higher total throughput compared to the baseline methods. This highlights the effectiveness of the {\LLMhRIC} framework in optimizing resource allocation and improving network performance. Furthermore, the decaying $w$ is critical for performance. A fixed $w=0$ leads to training failure from an abrupt policy shift, while a fixed $w=0.9$ causes performance to stagnate by suppressing the agent's exploration. In contrast, the proposed decaying $w$ ensures a smooth transition from guided exploration to self-directed optimization, which is necessary for stable convergence to the highest-performing policy.

\begin{figure}[t]
	\centering
	\includegraphics[width=0.8\linewidth]{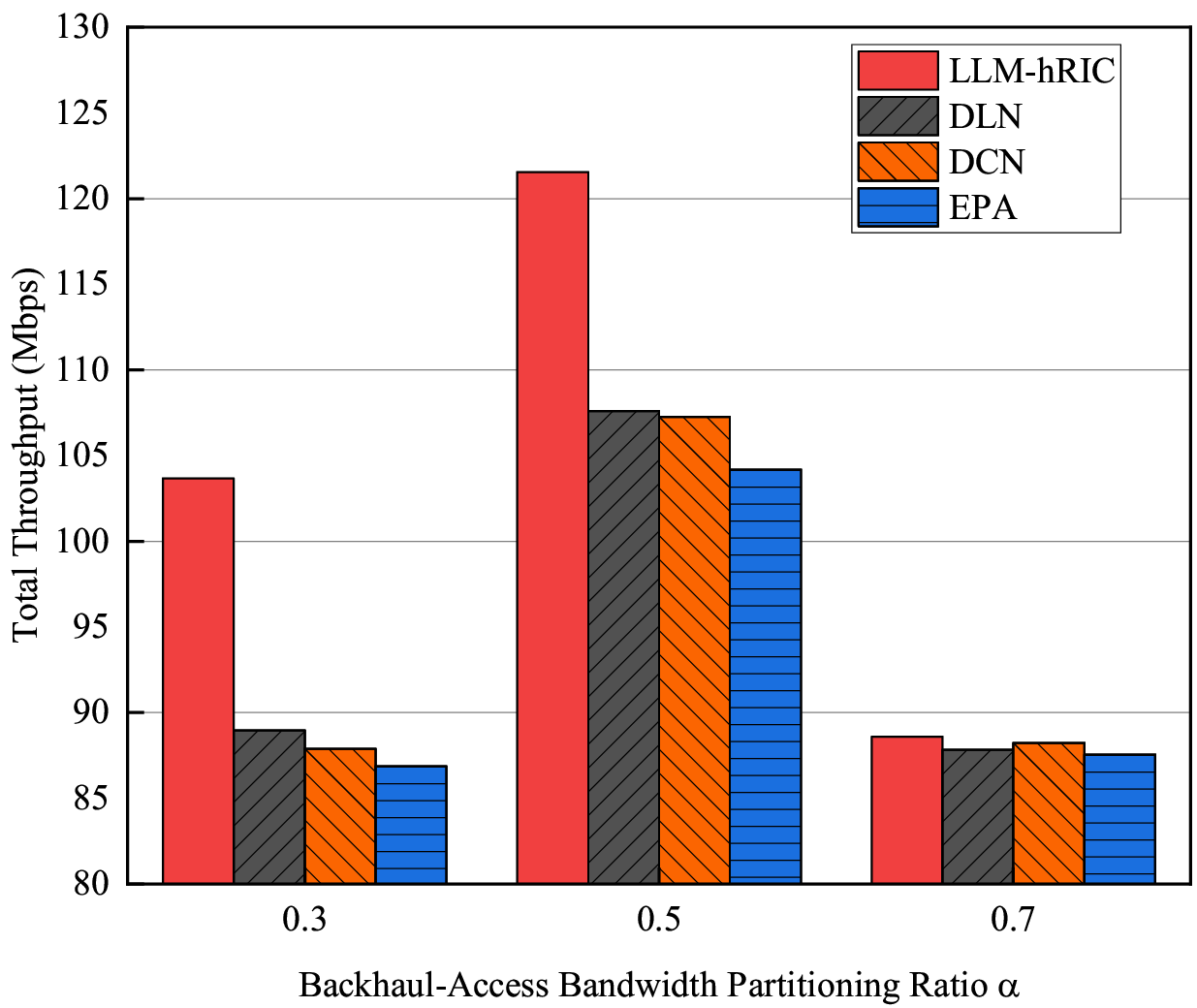}
	\caption{Average total throughput of the proposed {\LLMhRIC} and baseline methods according to backhaul-access bandwidth partitioning ratio $\alpha$ over 50 test examples.}
	\label{fig:Bandwidth}
\end{figure}
Fig. \ref{fig:Bandwidth} illustrates the effect of the backhaul-access bandwidth partitioning ratio on the total throughput. The result shows that the throughput initially increases and then decreases as the allocation ratio $\alpha$ increases. This is because the performance bottleneck exists at the backhaul link for a low $\alpha$, while it is primarily in the access link for large $\alpha$. The proposed {\LLMhRIC} framework consistently outperforms baseline methods. This is due to the high-quality initial policies, generated by \ac{LLM}, which enhances exploration efficiency and facilitates better policy discovery by \ac{RL}. The \ac{LLM} leverage detailed, and global information to improve coordination among \acp{MBS}, resulting in a more efficient resource allocation and higher overall network performance.

\vspace{-1mm}
\section{Future Challenges} \label{sec:future challenges}
\begin{figure*}[h]
	\centering
	\includegraphics[scale=0.75]{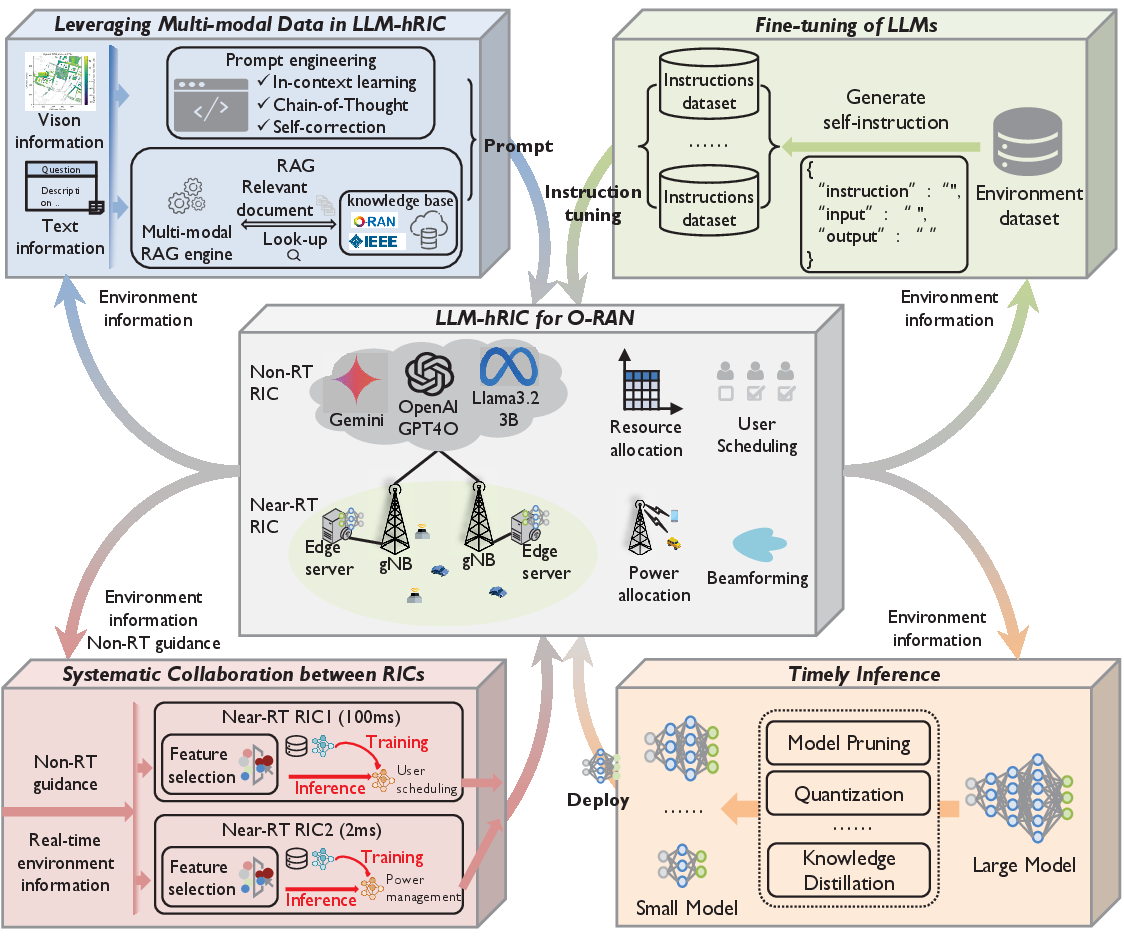}
	\caption{Future challenges of the {\LLMhRIC} for \ac{O-RAN}.}
	\label{fig:challenges}
\end{figure*}

In this section, we discusses the future challenges of {\LLMhRIC} for \ac{O-RAN}. As shown in Fig. \ref{fig:challenges}, key challenges specific to {\LLMhRIC} include leveraging multi-modal data, fine-tuning \acp{LLM} for guidance generation, enabling systematic collaboration between \acp{RIC} in {\LLMhRIC}, and achieving timely inference. Each topic is discussed in detailed below.

\vspace{-2mm}
\subsection{Leveraging Multi-Model Data in {\LLMhRIC}}
In the {\LLMhRIC}, the multi-modal \acp{LLM} can reason across various data types. Multi-modal data (e.g., text, signals, images, and videos) provides detailed environment insights, enhancing the ability of the \ac{non-RT RIC} to generate more accurate and context-aware guidance for \ac{near-RT RIC}. 
Moreover, zero-shot and few-shot learning techniques, such as prompt engineering and \ac{RAG}, have been explored to enhance the performance of \acp{LLM} in new environments.  
Despite their potential, significant challenges remain: 1) prompt engineering lacks robust support for multi-modal inputs in the wireless communication domain, necessitating new methods to handle such data for generating high-quality guidance. 
2) \ac{RAG} performance relies heavily on the retrieval engine. Designing an effective multi-modal \ac{RAG} engine for {\LLMhRIC} in the wireless communication domain is critical to improve \ac{LLM} reasoning and guidance generation.

\subsection{Fine-Tuning of \acp{LLM}}
Current \acp{LLM} are trained for general domains but not optimized for wireless communication tasks, limiting guidance quality in {\LLMhRIC}. 
While prompt engineering can partially bridge this gap, its effectiveness is limited by the quality of examples and input token window size.
Moreover, the guidance from a general-purpose \ac{LLM} can sometimes be unstable, making fine-tuning important for adapting to environmental changes and ensuring reliability.
Fine-tuning \acp{LLM} with offline, domain-specific data is essential to enhance guidance reliability in the \ac{non-RT RIC}.
However, three challenges remain: 1) Collecting or effectively utilizing the limited available data on wireless communication is challenging. 2) Generating high-quality instruction data for fine-tuning is essential. Automated or semi-automated methods for generating diverse, domain-relevant instruction data must be developed to improve the guidance quality of fine-tuned \acp{LLM}. 3) A privacy-preserving architecture should be considered such as the combination of federated learning (FL) with differential privacy (DP). The DP adds controlled noise to the shared model updates in FL for a mathematical guarantee that private information on the training data is not leaked. 

\subsection{Systematic Collaboration between \acp{RIC}}
In {\LLMhRIC}, the \acp{LLM}-empowered \ac{non-RT RIC} generates guidance for multiple \acp{near-RT RIC}, each responsible for diverse tasks. A systematic collaboration framework is needed to ensure smooth interaction between the \acp{LLM} and lightweight AI models.
Several challenges need to be addressed. 1) \acp{non-RT RIC} operates on a large time scale, while \acp{near-RT RIC} can operate on different small time scales. This temporal mismatch complicates the synchronization between the guidance generation in the \ac{non-RT RIC} and the real-time decision-making in the \ac{near-RT RIC}. 2) \acp{near-RT RIC} perform a variety of tasks (e.g., power allocation, \ac{PRB} allocation, and user association). Each task requires specific guidance tailored to its unique requirements, making it crucial to develop a mechanism for providing task-aware, customized guidance for different \acp{near-RT RIC}.

\subsection{Timely Inference}
Modern \acp{LLM}, such as GPT-4 with 1.7 trillion parameters, require significant computational resources and are often cloud-deployed, causing latency issues. This makes deploying an \ac{LLM} in the \ac{non-RT RIC} challenging, as some \ac{non-RT RIC} require latency near 1s. 
Technologies like model pruning, quantization, and knowledge distillation can reduce model size. Specifically, the quantization technology reduces the precision of model parameters to lower computational overhead, while pruning and distillation create smaller, optimized \acp{LLM}. These methods compress models at varying ratios but often trade off performance for speed. Balancing model size, inference speed, and accuracy is critical to ensuring efficient deployment of \acp{LLM} in the {\LLMhRIC} framework.

\section{Conclusions}
This paper proposes the {\LLMhRIC} framework for \ac{O-RAN}. 
In this framework, 
the \ac{LLM}-empowered \ac{non-RT RIC} provides the strategic guidance, while the \ac{RL}-empowered \ac{near-RT RIC} explores the guidance with local observation to facilitate near-RT decision-making. The superior performance demonstrates the effectiveness of combining \ac{LLM}-empowered \ac{non-RT RIC} and \ac{RL}-empowered \ac{near-RT RIC} to tackle the dynamic and complex resource management.
Meanwhile, this work highlights the potential of integrating global reasoning and local responsiveness, establishing the {\LLMhRIC} framework as a foundation for scalable, adaptive, and high-performance solutions in future \ac{O-RAN} systems. 
For instance, the framework is extensible to advanced paradigms like the intent-based networking, where the LLM can translate an administrator's high-level and multi-objective intent into strategic guidance. This guidance is then executed by RL agents with real-time and tactical adaptations, showcasing a powerful synergy for future network automation.
Additionally, we identify key challenges for {\LLMhRIC}, as discussed in Section `Future Challenges',
% Section~\ref{sec:future challenges}, 
to guide future research and development.

% \bibliographystyle{IEEEtran}
% \bibliography{Bib/IEEEabrv,Bib/ISCGroup,Bib/StringDefinitions,Bib/bib_LLM}

\vspace{-10mm}
\begin{IEEEbiographynophoto}
    {Lingyan Bao} [S'21] (lingyan@yonsei.ac.kr) is currently a Ph.D. student in the School of Electrical and Electronic Engineering, Seoul, Yonsei University, Seoul, Korea.
\end{IEEEbiographynophoto}

\vspace{-10mm}

\begin{IEEEbiographynophoto}
    {Sinwoong Yun} [S'19] (syun4656@etri.re.kr) is a Reseacher at Electronics and Telecommunications Research Institute, Daejeon, Korea.
\end{IEEEbiographynophoto}

\vspace{-10mm}

\begin{IEEEbiographynophoto}
    {Jemin Lee} [S`06-M`11-SM'24] (jemin.lee@yonsei.ac.kr) is an Associate Professor at the School of Electrical and Electronic Engineering, Yonsei University, Seoul, Korea.
\end{IEEEbiographynophoto}

\vspace{-10mm}

\begin{IEEEbiographynophoto}
    {Tony Q. S. Quek} [S'98-M'08-SM'12-F'18] (tonyquek@sutd.edu.sg) is the Cheng Tsang Man Chair Professor with Singapore University of Technology and Design (SUTD) and ST Engineering Distinguished Professor.
\end{IEEEbiographynophoto}
\end{document}